\documentclass[preprint,showpacs,pra]{revtex4}
\usepackage{graphicx}
\begin{document}
%
\title{ K-Rb Fermi-Bose mixtures: vortical states and sag }
\author{D. M. Jezek}
\affiliation{Departamento de F\'{\i}sica, Facultad de Ciencias Exactas
y Naturales, \\
Universidad de Buenos Aires, RA-1428 Buenos Aires, and \\
Consejo Nacional de Investigaciones Cient\'{\i}ficas y T\'ecnicas,
Argentina}
\author{M. Barranco, M. Guilleumas, R. Mayol, and M. Pi}
\affiliation{Departament d'Estructura i Constituents de la Mat\`eria,
Facultat de F\'{\i}sica, \\
Universitat de Barcelona, E-08028 Barcelona, Spain}
%
%
%
\begin{abstract}
We study a confined mixture of bosons and fermions in the
quantal degeneracy regime with attractive boson-fermion
interaction. We discuss the effect that the presence of vortical
states and the displacement of the trapping potentials may have on
mixtures near collapse, and investigate the phase
stability diagram of the K-Rb mixture in the mean-field approximation
supposing in one case that 
the trapping potentials felt by bosons and fermions are shifted
from each other, as it happens in the presence of a gravitational sag, and
in another case, assuming that the Bose condensate sustains a
vortex state. In both cases, we have obtained an analytical expression 
for the fermion effective potential 
when the Bose condensate is in the Thomas-Fermi regime,
that can be used to determine the maxima of the fermionic density. We have
numerically checked that the values one obtains for the location
of these maxima using the analytical formulas remain valid up to
the critical  boson and fermion numbers, above which the mixture
collapses.

\end{abstract}
\pacs{03.75.-b, 03.75.Mn, 03.75.Lm, 32.80.Pj}
\maketitle
%
\section{Introduction}

Recent experiments on degenerate Fermi-Bose mixtures have opened
the possibility of studying  in a direct way the effect of quantum
statistics in Bose-Einstein condensates (BECs). In practice, the
direct evaporative cooling techniques  used to obtain  BEC are not
applicable in a Fermi gas, as  the Pauli exclusion principle
forbids s-wave collisions between fermions. This difficulty has
been overcome using a gas of bosons  as a coolant, which has given
further relevance to the study of these mixtures.

One of the systems studied in recent experiments  is the
$^6$Li-$^7$Li mixture \cite{tr01,sc01}.  It is characterized by
having a positive boson-fermion scattering length  an order of
magnitude larger than the boson-boson one \cite{sc01}. Due to this
repulsive boson-fermion interaction,  the   system does not
exhibit a large overlap between the two species  \cite{ro02}.
Moreover, this mixture may undergo a two-component  separation
\cite{ro02,ak02}. Both facts conspire against having the fermions
well inside the boson cloud, which is desirable to obtain an
efficient sympathetic cooling. In contrast, this desirable overlap
can be achieved in $^{40}$K-$^{87}$Rb mixtures due to the large
attractive boson-fermion interaction. This mixture has been
recently obtained \cite{fe02,roa02,mo03}, and it
has been shown that if the particle numbers are above some
critical values the system collapses \cite{mo03}.

From a theoretical point of view, a fairly amount of work has been
devoted to the study of boson-fermion mixtures \cite{ro02,mo98,vi00}. In
particular, a systematic study of the structure of
binary mixtures has been performed in Ref. \cite{ro02},
where all possible sign combinations of scattering lengths between
boson-boson and boson-fermion s-wave interactions have been
discussed.
The purpose of our work is, firstly,  to present  an
analysis of the location of the minima of the effective potential
felt by fermions submitted to a large number of condensate
bosons when the boson-boson interaction is repulsive, and secondly,
to extend this study  to systems with a
large number of fermions with an attractive boson-fermion interaction.

An essential characteristic  of superfluid systems is the
occurrence of quantized vortices \cite{don91}. Quantized vortices
in BECs were first produced experimentally by Matthews et al.
\cite{mat99}, and have been studied since  in  detail
(see Refs. \cite{fet01,pet02,pit03} and Refs. therein). It may
thus be interesting to see what is the physical appearance of such
vortices, arising in the bosonic component, in the presence of a
fermionic cloud which is in the normal (non superfluid) but
quantum degenerate phase. This is another aim of our work,
which bears some similarities with the description of quantized
vortices in $^3$He-$^4$He nanodroplets recently
addressed \cite{may01}.

This  work is organized as follows.
In Sec. II we describe the mean-field model we have used.  In Sec. 
III we present an analysis of the effective potential felt by the
fermions when they are inside a large condensate that exhibits a
boson-boson repulsive interaction, and the number of fermions ($N_F$) is
much smaller than the number of bosons ($N_B$).
We shall consider the case in
which the boson and fermion  external confining potentials 
are displaced from each other, which may  be caused by
gravity for instance, and also when bosons are in a vortical state. In
Sec. IV we present the results obtained by solving the
mean-field coupled equations  for different 
$N_B$ and $N_F$ values up to the critical values where
collapse of the mixture occurs. 
A summary is given in Sec. V. Finally, in the
Appendix we derive some expressions  using a scaling
transformation which, on the one hand, constitute a generalization
of the virial theorem, and on the other hand are especially useful
for testing the accuracy of the numerical procedure.

\section{The mean-field model}

We consider a mixture of a Bose condensate (B) and a degenerate
Fermi gas (F) at zero temperature confined in an axially symmetric
harmonic trap. Assuming that the minimum of the trapping potential
felt by each species may be displaced in the $z$ direction in a
value $d_i$ (i=B,F), the confining potentials in cylindrical
coordinates 
are 
\begin{equation}
V_B=\frac{1}{2}\,M_B\,\left[\omega_{rB}^2\,r^2 +
\omega_{zB}^2\,(z-d_B)^2\right]
\end{equation}
and
\begin{equation}
 V_F=\frac{1}{2}\,M_F\,\left[\omega_{rF}^2\,r^2 +
\omega_{zF}^2\,(z-d_F)^2\right] \;\;\; ,
\end{equation}
where $\omega_{rB}$, $\omega_{zB}$ and $\omega_{rF}$,  $\omega_{zF}$
are the  trapping radial (axial) angular frequencies
for bosons and fermions, respectively. $M_B$ and $M_F$ are the
corresponding masses.

Since the number of fermions we consider in the
numerical calculations is fairly large,
the fermionic kinetic energy density can be written in the
Thomas-Fermi-Weizs\"acker (TFW) approximation as a function of the
local fermion density $n_F$ and its gradients. For fully
polarized spin $1/2$ fermions, it reads:
\begin{equation}
 \tau_F( \vec{r}\,)=
 \frac{3}{5} (6 \pi^2)^{2/3} n_F^{5/3}+ \beta \frac{(\nabla n_F)^2}
 {n_F}
\label{tau}
\end{equation}
and the fermionic kinetic energy is 
\begin{equation}
T_F=  \frac{\hbar^2}{2 M_F} \int d \vec{r}\, \tau_F(\vec{r}\,)\,.
\label{ekin}
\end{equation}
The value of the $\beta$ coefficient in the Weizs\"acker term is
fixed to 1/18. This term contributes little to the kinetic
energy, and it is usually neglected \cite{ro02}. 
However, its inclusion\cite{ca03}
has some advantages, see below. We refer the interested reader to Ref.
\cite{pi88} for a discussion on the accuracy of the TFW
approximation (see also Refs. therein).  

Neglecting all p-wave interactions, the energy density functional
for the boson-fermion mixture at zero temperature has the form
\begin{equation}
 {\cal E}( \vec{r}\,)=\frac{ \hbar^2 }{2 M_B}  |\nabla \Psi |^2 +
V_B \,n_B + \frac{1}{2} g_{BB}\, n_B^2 + g_{BF} \,n_F\, n_B +
 \frac{\hbar^2}{2 M_F} \tau_F + V_F \,n_F \;\;\; ,
\label{ed}
\end{equation}
where $ n_B= |\Psi|^2$ is the local boson atomic density. The
boson-boson and boson-fermion coupling constants $g_{BB}$ and
$g_{BF}$ are written in terms of the $s$-wave scattering lengths
$a_B$ and  $a_{BF}$  as $g_{BB}=4\pi\,a_{B}\hbar^2/M_B$ and
$g_{BF}=4\pi\,a_{BF}\hbar^2/M_{BF}$, respectively. We have defined
$M_{BF} \equiv 2 M_B M_F / (M_B + M_F)$.

When bosons sustain a quantized vortex line along the
$z$-axis, the condensate wave function can be written as $ \Psi=
|\Psi| e^{i \,m \,\phi }$, where $m= 1,2,3 \ldots$ is the
circulation number, yielding for the kinetic energy density
\begin{equation}
\frac{ \hbar^2 }{2 M_B}  |\nabla \Psi |^2 =
\frac{ \hbar^2 }{2 M_B}  (\nabla |\Psi |)^2 +
 \frac{ \hbar^2 m^2}{2 M_B}  \frac{n_B}{r^2} \,.
\label{centri}
\end{equation}
We have considered only singly quantized vortices, that is $m=1$. 
If a vortex is present in the condensate,
bosons flow around the vortex core with quantized circulation,
which yields the centrifugal term in the kinetic energy. 
We assume that the Fermi component is not superfluid,
and consider that it is in a stationary state.
This situation could be achieved
experimentally waiting enough time after the generation of the
vortex in the condensate, to let
the drag force between bosons and fermions to dissipate.

Variation of ${\cal E}$ with respect to $\Psi$ and $n_F$
keeping $N_B$ and $N_F$ fixed yields
the following coupled Euler-Lagrange (EL) equations
\begin{equation}
 \left (-\frac{ \hbar^2 \nabla^2}{2 M_B}+ V_B  +
 \frac{ \hbar^2 m^2}{2 M_B}  \frac{1}{r^2}
+ g_{BB}\, n_B + g_{BF}\, n_F
 \right ) \, \Psi = \mu_B \Psi
\label{gp}
\end{equation}
\begin{equation}
 \frac{ \hbar^2 }{2 M_F}
\left[(6 \pi^2)^{2/3} n_F^{5/3} + \beta \frac{(\nabla n_F)^2}{n_F}
- 2 \beta \Delta n_F \right] + V_F\, n_F +  g_{BF}\, n_B\, n_F
 \,  = \mu_F \,n_F                \;\;\; ,
\label{tf}
\end{equation}
where $ \mu_B$ and $ \mu_F$ are the boson and fermion chemical
potentials, respectively. Then, the ground state ($m=0$) or a
vortical state ($m=1$) are obtained by solving the
Gross-Pitaevskii (GP) equation for bosons [Eq. (\ref{gp})] coupled to
the Thomas-Fermi-Weizs\"acker equation for fermions [Eq. (\ref{tf})].

The inclusion of the Weizs\"acker term in the energy density has
the major advantage that it yields a EL equation for $n_F$ that is
well behaved everywhere, avoiding the classical turning
point problem when this term is neglected (Thomas-Fermi approximation).
Moreover, solving Eq.
(\ref{tf}) is not more complicated than solving the GP equation.
This can be readily seen writing the later in terms of $n_B$:
\begin{equation}
 \frac{ \hbar^2 }{2 M_B} \left[ \frac{1}{4} \frac{(\nabla n_B)^2}{n_B}
-\frac{1}{2}  \Delta n_B \right] + V_B \,n_B + \frac{ \hbar^2
m^2}{2 M_B}  \frac{n_B}{r^2} + g_{BB}\, n_B^2 + g_{BF}\, n_B\, n_F
\,  = \mu_B \,n_B                \;\;\; , \label{gpnb}
\end{equation}
which is formally equivalent to Eq. (\ref{tf}). We have
discretized these equations using 9-point formulas and have solved
them on a 2D ($r,z$) mesh using a sufficiently large box. The
results have been tested for different sizes of the spatial steps
(we have mostly used $\Delta r= \Delta z \sim 0.1 \mu m$).
We have employed the imaginary time
method to find the solution of these coupled equations written as
imaginary time diffusion equations \cite{bar03}.  After
every imaginary time step, the densities are normalized to the
corresponding particle numbers. To start the iteration procedure we
have used positive random numbers to build both normalized
densities. This avoids to introduce any bias in the
final results. We have also checked that the solutions fulfill
the generalized virial theorem deduced in the Appendix.

\section{Effective fermion potential}
Before we present the numerical results obtained by solving  Eqs.
(\ref{tf}) and (\ref{gpnb}), it is useful to analyze  the features
of the effective potential felt by
a small number of  fermions in the presence of a boson condensate,
paying special attention to the location of its minima.
 We are interested in studying  the cases in which either the 
 condensate hosts a vortex line
or there exists a displacement between the  minima of the 
boson and fermion external potentials. The later situation is
routinelly met in the experiments -a gravitational sag in
the $z$ direction-, in which case the displacement is $ g
(1/\omega_{zF}^2 - 1/\omega_{zB}^2 ) $, being $g$ the acceleration
of gravity.

An interesting issue is that the analytic expressions we obtain
in this Sec. for the location of the minimum of the effective potentials
remain valid, as we will show in the next Sec.,  
when the number of fermions and bosons are similar. 
These locations coincide with the positions of the maxima
of the fermion density and are relevant because
the collapse of the mixture originates precisely around them.
This will be discussed in the next Sec.

The effective fermion potential has  two main terms,
the external potential and  the mean-field term arising from the 
interaction with the boson condensate, which is proportional to $n_B$.
We consider a large condensate in the Thomas-Fermi (TF) regime 
with positive scattering length and we assume that
its density profile is  not affected by the fermion presence. To obtain 
$n_B$  we may thus use  Eq. (\ref{gp}) with $n_F=0$ and neglect
the  first kinetic energy term.

\subsection{Effect of a shift in the minimum of the external potentials}

For simplicity,
we restrict ourselves to the vortex-free case $m=0$. Without loss
of generality, the origin of coordinates can be fixed at the
minimum of the boson trapping potential, and we will assume
that the displacement is only in the $z$ direction. Thus, the
effective trapping potential experienced by the fermions is
\begin{equation}
 V_{d}^{eff} =  \frac{1}{2}\,M_F\,\left[\omega_{rF}^2\,r^2 +
\omega_{zF}^2\,(z-d_z)^2\right] + g_{BF}\, n_B \;\;\; ,
\label{veffd}
\end{equation}
where $d_z=d_F-d_B$ is
the $z$ displacement between the trapping potential
centers. Assuming that the density profile of the
condensate is not affected by the interaction with fermions,
the number of bosons is large,  and that the interaction between
bosons is repulsive ($ g_{BB} >0$), one obtains
from Eq. (\ref{gp}) setting $m=0$ the
boson density profile that corresponds to the uncoupled TF density
\begin{equation}
n_B =
\frac{1}{g_{BB}}\left[ \mu_B - \frac{1}{2}\,M_B\, (\omega_{rB}^2\,
r^2 + \omega_{zB}^2\,z^2 ) \, \right]\,\Theta\left[\mu_B -
\frac{1}{2}\,M_B\,(\omega_{rB}^2\,
 r^2 + \omega_{zB}^2\, z^2) \right] 
\label{denb}
\end{equation}
with $\Theta(x)=1$ if $x>0$ and zero otherwise. The TF condensate
boundary is given by the ellipsoid $arg(\Theta)=0$. 
For an axially symmetric trap it yields the well-known TF
radius of the condensate
$R_i=\sqrt{2 \mu_B/(M_B \, \omega_{iB}^2)}$ with $i=r,z$
in the radial and axial direction, respectively.
Replacing $n_B $ into Eq.\ (\ref{veffd}) and defining the 
dimensionless parameter
\begin{equation}
 \gamma_i \equiv 1-\frac{a_{BF} M_{B}^2 \,\omega_{iB}^2}
{ a_{B} M_{BF} M_{F} \,\omega_{iF}^2}
\label{eqgamma}
\end{equation}
with $i=r,z$, it follows that the effective fermion potential
inside the Bose condensate -where  $n_B$ [Eq. (\ref{denb})] is
positive- is
\begin{equation}
 V_{d}^{eff} =  \frac{1}{2}\,M_F\,\left[ \gamma_r\, \omega_{rF}^2\, r^2
 + \gamma_z \,\omega_{zF}^2 (z-d_z/\gamma_z)^2\right]
 + \frac{1}{2}\,M_F \, \omega_{zF}^2 \, d_z^2 ( 1 - \gamma_z^{-1})
+  \frac{g_{BF}}{g_{BB}} \mu_B
\;\;\; ,
\label{veff2}
\end{equation}
whereas outside the condensate
\begin{equation}
 V_{d}^{eff} =  \frac{1}{2}\,M_F\,\left[\omega_{rF}^2\,r^2 +
\omega_{zF}^2\,(z-d_z)^2\right] \label{veff3} \,.
\end{equation}
This effective potential  can be viewed as having a renormalized
frequency inside the  boson condensate, a feature already
discussed in Ref. \cite{vi00} for a mixture with concentric
external potentials.
This model has been called the double-parabola model by
Vichi et al. \cite{vi00}, and it has also been used by Capuzzi and
Hern\'andez \cite{ca01}. However, in these works the possibility of a
displacement between  the minima of  the potentials has not been
considered.

The extremum of $V_{d}^{eff}$ inside the Bose condensate is attained at
the point
\begin{equation}
 \vec{d}'=(d'_x, d'_y, d'_z)=(0,0, \gamma_z^{-1} d_z)
\;\;\;  .
\label{rm}
\end{equation}
 Note that
if the shift between the external potentials is $\vec{d} =(d_x,
d_y, d_z)$, the effective potential has its extremum at $
\vec{d}' =( \gamma_x^{-1} d_x,\, \gamma_y^{-1} d_y,\,
\gamma_z^{-1} d_z)$. Depending on  the signs of $\gamma_i$, $
\vec{d}'$ may correspond to a maximum, a minimum or a saddle
point.

If the boson-fermion interaction is attractive ($g_{BF}<0$) as for
$^{40}$K-$^{87}$Rb mixtures,  $ \gamma_i >1 $ and the
point $ \vec{d}'$ is a minimum.  Depending on the positions of
$\vec{d}$ and $\vec{d}'$ with  respect to the ellipsoid $arg
(\Theta)=0$, with the argument of $\Theta$ taken from Eq.\
(\ref{denb}), there are three possibilities: a) If $\vec{d}$ is
 inside this ellipsoid, fermions view only one minimum at
$\vec{d}'$; b) If $\vec{d}$  is outside and  $\vec{d}'$ is inside,
there are two minima at $\vec{d}$ and $\vec{d}'$;
c) If both points are outside the
ellipsoid,  there is only one minimum at $\vec{d}$.

\subsection{Vortices}
We consider now the effect of a  quantized boson
vortex line in the fermion distribution, without
gravitational sag. In this case, the effective
fermion potential is
\begin{equation}
 V_{v}^{eff} =
 V_F  + g_{BF} \,n_B^v =
   \frac{1}{2}\,M_F\,\left(\omega_{rF}^2\,r^2 +
\omega_{zF}^2\, z^2\right) + g_{BF}\, n_B^v \;\;\; ,
\label{veffv}
\end{equation}
where $n_B^v$ is the boson density hosting a vortex line along the
$z$ axis.  This density is zero at $r=0$, and reaches its maximum
value on a circle of radius $r_0$ around that axis.  In the TF
approximation, $n_B^v$ can be derived using Eq. (\ref{gp}) keeping
the centrifugal term proportional to $m$,
 and may be  approximately written 
as \cite{fet01}
\begin{equation}
n_B^v=  \frac{1}{g_{BB}}\left[ \mu_B - \frac{ \hbar^2 m^2}{2 M_B}
\frac{1}{r^2} - V_B \, \right]\,
\Theta\left[\mu_B - \frac{ \hbar^2 m^2}{2 M_B}  \frac{1}{r^2}
-  V_B  \right]
\end{equation}
with $r_{0} = \sqrt{\hbar \, m /( M_B \, \omega_{rB}})$.
Using $n_B^v$  we calculate the mimimum of $ V_{v}^{eff}$
inside the condensate, which is located at $z'=0$ and
\begin{equation}
 r' = (1-\gamma_r^{-1})^{1/4} \;
  \sqrt{\frac{ \hbar \, m}{ M_B \, \omega_{rB}}}
= (1-\gamma_r^{-1})^{1/4} \,   r_0  \;\;\; .
\label{rmin}
\end{equation}
Thus, the minimum is attained at a circle of radius $r'$.
We will see that for the $^{40}$K-$^{87}$Rb mixture
$r'$ is very close to the radius at
which the boson density has its maximum.

\section{Numerical results}

The system under consideration is a confined $^{40}$K-$^{87}$Rb mixture.
We have assumed spherically symmetric traps for bosons and fermions,
with trap frequencies $\omega_B = 2 \pi \times 100$ Hz and 
$\omega_F = \sqrt{M_B/M_F}\; \omega_B$. When a displacement $d_z$
is introduced between the  minima of the external potentials,
the mixture has only axial symmetry around the $z$ axis. 
We have numerically solved Eqs. (\ref{tf}) and (\ref{gpnb})
using the set of scattering lengths reported by Modugno et al.
\cite{mo03}, namely  $a_B= 98.98\, a_0 $, and $a_{BF}= -395\, a_0$,
being $a_0$  the Bohr radius.
The dimensionless parameters introduced in Eq. (\ref{eqgamma})
are $\gamma_z^{-1} = \gamma_r^{-1}=0.136$, and
$(1- \gamma_r^{-1}) ^{1/4} = 0.964$ [Eq. (\ref{rmin})].

\begin{figure}
\centerline{\includegraphics[width=16cm,angle=-90]{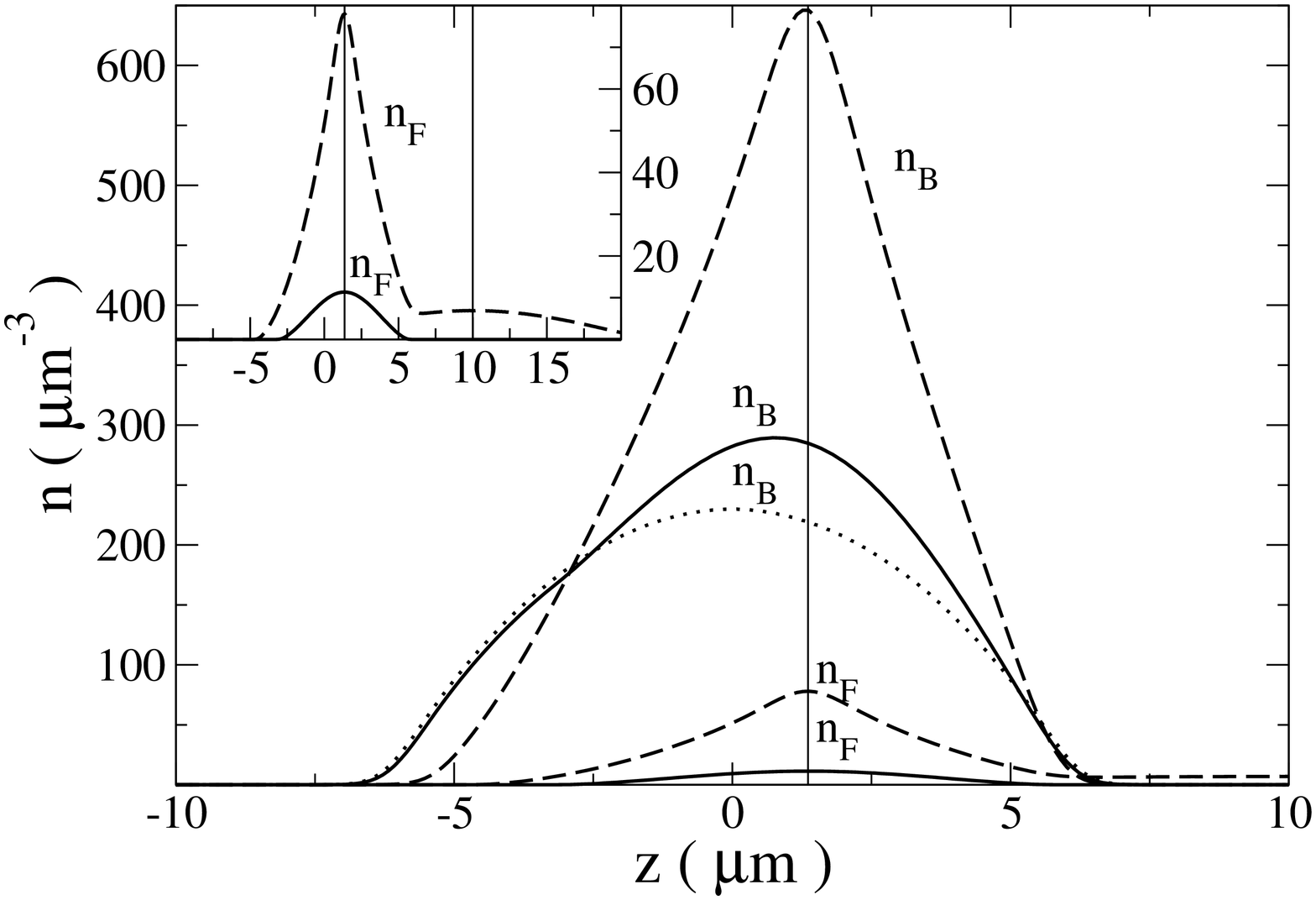}}
\caption[]{
Boson and fermion density profiles of $^{40}$K-$^{87}$Rb mixtures
as a function of $z$ for
a $ d_z = 10 \,\mu$m displacement. The different lines correspond to 
the profiles for $N_B= 10^5$  and several $N_F$ values:
$N_F= 0$ (dotted line);
$N_F= 10^3$ (solid lines);
$N_F= 2.5 \times 10^4$ (dashed lines).
The inset shows a magnified view of the fermion density distributions.
}
\label{fig1}
\end{figure}

We display in Fig. \ref{fig1} several  boson and fermion density profiles
as a function of $z$, considering an arbitrary displacement
$d_z = 10\, \mu$m. They all correspond to $N_B= 10^5$, but for  three 
different fermion numbers:
$N_F=0$ , $N_F=10^3$, and $N_F= 2.5 \times 10^4$.
The attractive boson-fermion interaction produces
an enhancement of the density of both species in the 
overlap volume. However, this effect is reduced with respect to the
concentric case due to $z$ displacement\cite{mo03}.

Using the TF approximation, the radius
of the Bose condensate can be estimated as
$R_B=(15 N_B a_B /a_{HO})^{1/5}a_{HO}$,
with $a_{HO}=\sqrt{\hbar/{M_B \omega_B}}$ being the oscillator length
of the boson trap; this yields $R_B\sim\,6 \mu$m.
Following the analysis performed in Sec. III.A, and using
$\gamma_z^{-1} =0.136$, we find that  $d'_z=1.36 \mu$m $ < R_B$;
thus, within that model there are two minima
in the effective fermion potential. One is at
$ \vec{d}=10\,\mu$m $\hat z$, i.e., beyond the TF radius, and the other is
inside the boson cloud, at $ \vec{d}' =1.36\,\mu$m $ \hat{z}$.
It is interesting to note from Fig. \ref{fig1} that in all cases,
the maxima of the fermion density are precisely at $z=d'_z$ and $z=d_z$,
which we have displayed in the inset as vertical lines.

We thus see that for $N_F/N_B$ equal to $1 \%$ (solid line)
or even $25 \%$ (dashed line),
the maxima of the fermion density appear at the values obtained 
using the  model of Sec. III.A.
This can be understood by solving Eqs. (\ref{gp}) and (\ref{tf}) with
$m=0$ (i.e., no vortex), 
without neglecting the  effect of fermions
in the GP equation as it was done in
the simplified model of Sec. III, and
assuming that all the gradient terms in both  Eqs. can be
neglected. This assumption is justified for the number of bosons and
fermions we have used: in the case of bosons, this is what the TF 
approximation is about, and in the case of fermions, the gradient term
is a correction to the leading term proportional to $n_F^{5/3}$.
Using Eq.\ (\ref{gp}), we obtain the following expression for $n_B$:
\begin{equation}
 n_B= \frac{1}{g_{BB}} [\mu_B - V_B - g_{BF} \,n_F ] \;\;\; .
\label{deb}
\end{equation}
Replacing it into Eq.\ (\ref{tf}) and deriving the resulting
expression with respect to $z$ we get
\begin{equation}
\frac{1}{3} \frac{\hbar^2}{M_F} (6 \pi^2)^{2/3} n_F^{-1/3}\,
\frac{\partial n_F}{\partial z} - \frac{g_{BF}}{g_{BB}} M_B\,
 \omega_{zB}^2 z + M_F \,\omega_{zF}^2 (z- d_z) -
\frac{g_{BF}^2}{g_{BB}} \,
\frac{\partial n_F}{\partial z}=0  \;\;\;  .
\end{equation}
The extremum of the fermion density is found by setting
$\partial n_F / \partial z =0 $, with $ n_F \ne 0 $.
After some algebra we find that
the following equation has to be fulfilled:
\begin{equation}
M_F \,\omega_{zF}^2 ( \gamma_z z - d_z )=0 \;\;\; ,
\end{equation}
whose solution
 coincides precisely with the value $z=d'_z$ we have found in the previous
Sec.

Whereas the maximum of the fermion density is located
at $z=d'_z$ irrespective of the value of the dilution
($N_F/N_B$), it can
be seen from the numerical results that
the maximum of the boson density moves from $z=0$
towards $z=d'_z$  as $N_F$ increases. 
In fact, Fig. \ref{fig1} shows 
that even for a rather small $N_F$ value ($N_F=10^3,N_B= 10^5$),
the shape of the condensate 
differs from the  parabolic-type profile yielded by the TF
approximation for a fermion-free condensate,
Eq. (\ref{denb}).
It can be also appreciated in the inset that
for a dilution of $N_F/N_B=25\%$,
a fairly large number of
fermions remains without mixing (long tail in the fermion density
profile outside the Bose condensate).

\begin{figure}
\centerline{\includegraphics[width=16cm,angle=-90]{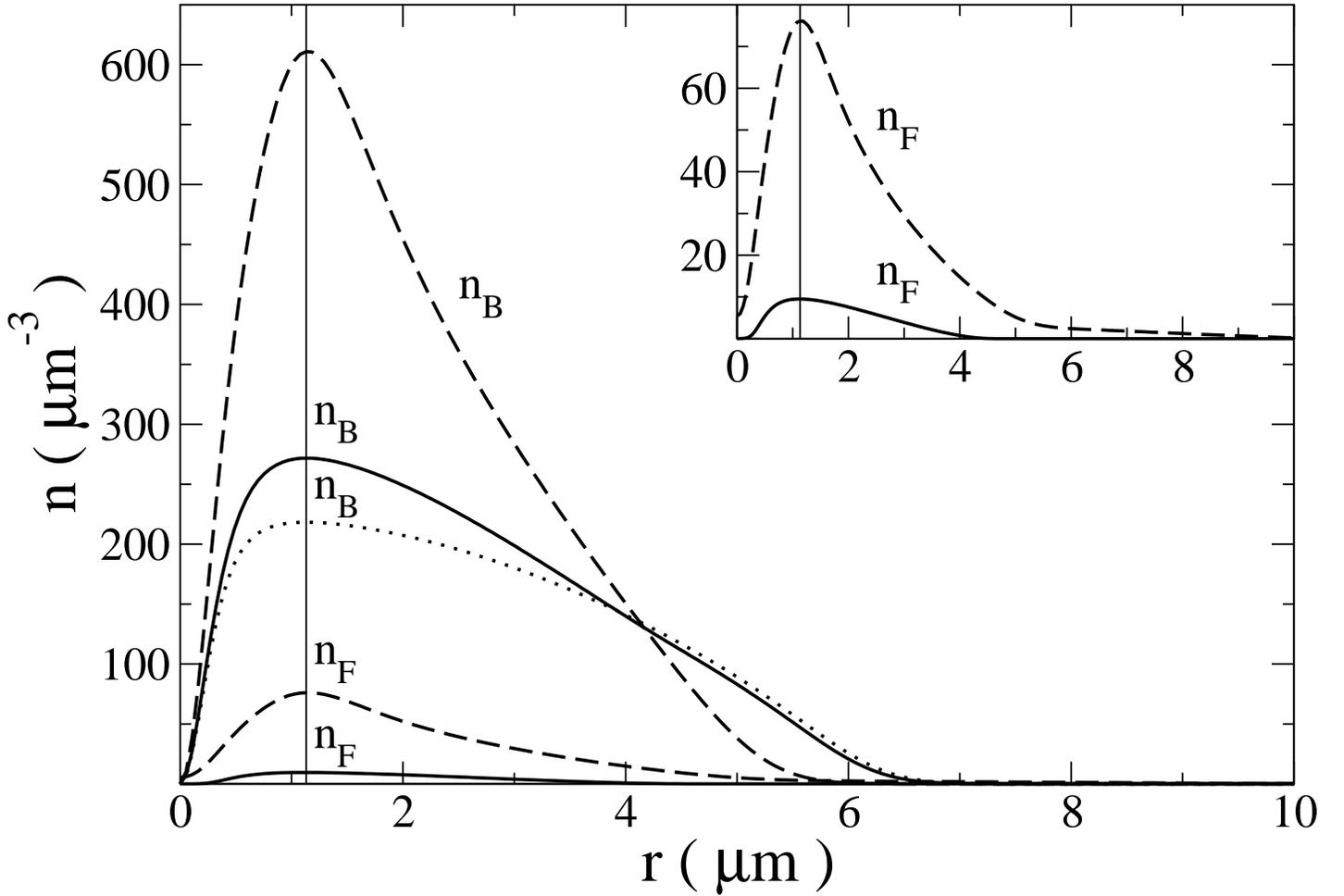}}
\caption[]{
Boson and fermion density profiles of $^{40}$K-$^{87}$Rb configurations
hosting a vortex line with $m=1$ without displacement of the trapping
potentials. In all cases, $N_B=10^5$. The different lines correspond to
$N_F= 0$ (dotted line);
$N_F = 10^3$ (solid lines); $ N_F= 1.5 \times 10^4$
(dashed lines).
The inset shows a magnified view of the fermion density distributions.
}
\label{fig2}
\end{figure}

We plot in Fig. \ref{fig2} several density profiles for configurations
hosting a singly quantized
vortex line in the case of no displacement between the
trapping potentials. We have taken
$N_B = 10^5 $, and have considered three different
fermion numbers, $ N_F=0$, $10^3$, and $1.5 \times 10^4$.
When $N_F=0$, the maximum of the bosonic density is located in the
$z=0$ plane,  in a circle of $r_0=1.13\, \mu$m  radius marked
with a vertical line in the graph.
Fermions are peaked around $ r'=0.96\, r_0 $,
as we have obtained using the simplified  model of Sec. III.B
with $(1- \gamma_r^{-1}) ^{1/4} = 0.964$.
 
It may be seen that in the presence of a quantized vortex,
the position of the maximum of boson and
fermion densities are very close 
irrespective  of the dilution $N_F/N_B$.
It is interesting to note that increasing $N_F$ at constant $N_B$, the  
fermion and boson density profiles  become sharper due to their mutual
attraction, eventually collapsing for a critical number of atoms.

Following a procedure analogous to that 
used before, we can justify the position of the maximum of the
fermion density in the presence of a vortex in a TF condensate.
From Eq.\ (\ref{gp}) we obtain
\begin{equation}
 n_B= \frac{1}{g_{BB}} \,
 [\mu_B - V_B - \frac{\hbar^2 m^2}{2 M_B} \frac{1}{r^2}
  - g_{BF}\, n_F ] \;\;\; .
\label{debv}
\end{equation}
Replacing it into Eq.\ (\ref{tf}) and deriving the resulting expression
with respect to $r$ we get
\begin{equation}
\frac{1}{3} \frac{\hbar^2}{M_F} (6 \pi^2)^{2/3} n_F^{-1/3}\,
\frac{\partial n_F}{\partial r} - \frac{g_{BF}}{g_{BB}} M_B\,
 \omega_{rB}^2 r + M_F \,\omega_{rF}^2 r
  + \frac{\hbar^2 m^2}{ M_B} \frac{1}{r^3}
  - \frac{g_{BF}^2}{g_{BB}}\,
\frac{\partial n_F}{\partial r}=0 \;\;\; .
\end{equation}
Once again, the extremum in the fermionic density is found by setting
$\partial n_F / \partial r =0 $ with $ n_F \ne 0 $, which
means that the following equation has to be fulfilled:
\begin{equation}
 - \frac{g_{BF}}{g_{BB}} M_B \,
 \omega_{rB}^2 r + M_F \,\omega_{rF}^2 r
  + \frac{\hbar^2 m^2}{ M_B} \frac{1}{r^3}
=0 \;\;\; ,
\end{equation}
whose solution
 again coincides  with $r=r'$ as found in the previous Sec. [Eq.
(\ref{rmin})].

Due to the attractive boson-fermion interaction, stable trapped
$^{40}$K-$^{87}$Rb mixtures may only have a limited number of
fermions and bosons. If the atom numbers increase above some
critical values $N_B^c$ and $N_F^c$, an instability 
occurs\cite{ro02}. It has been shown that the mean-field 
framework is able to reproduce the critical numbers for 
collapse\cite{mo03}. We have calculated the stability diagram of the
$^{40}$K-$^{87}$Rb mixture by solving the coupled mean-field
equations (\ref{tf}) and (\ref{gpnb}) for different values of $N_B$ 
and $N_F$. In our study, the instability signature is the failure 
of the numerical iterative process. In particular, the instability onset
appears as an indefinite growth of the maximum of the densities,
which triggers the collapse.

\begin{figure}
\centerline{\includegraphics[width=16cm,angle=-90]{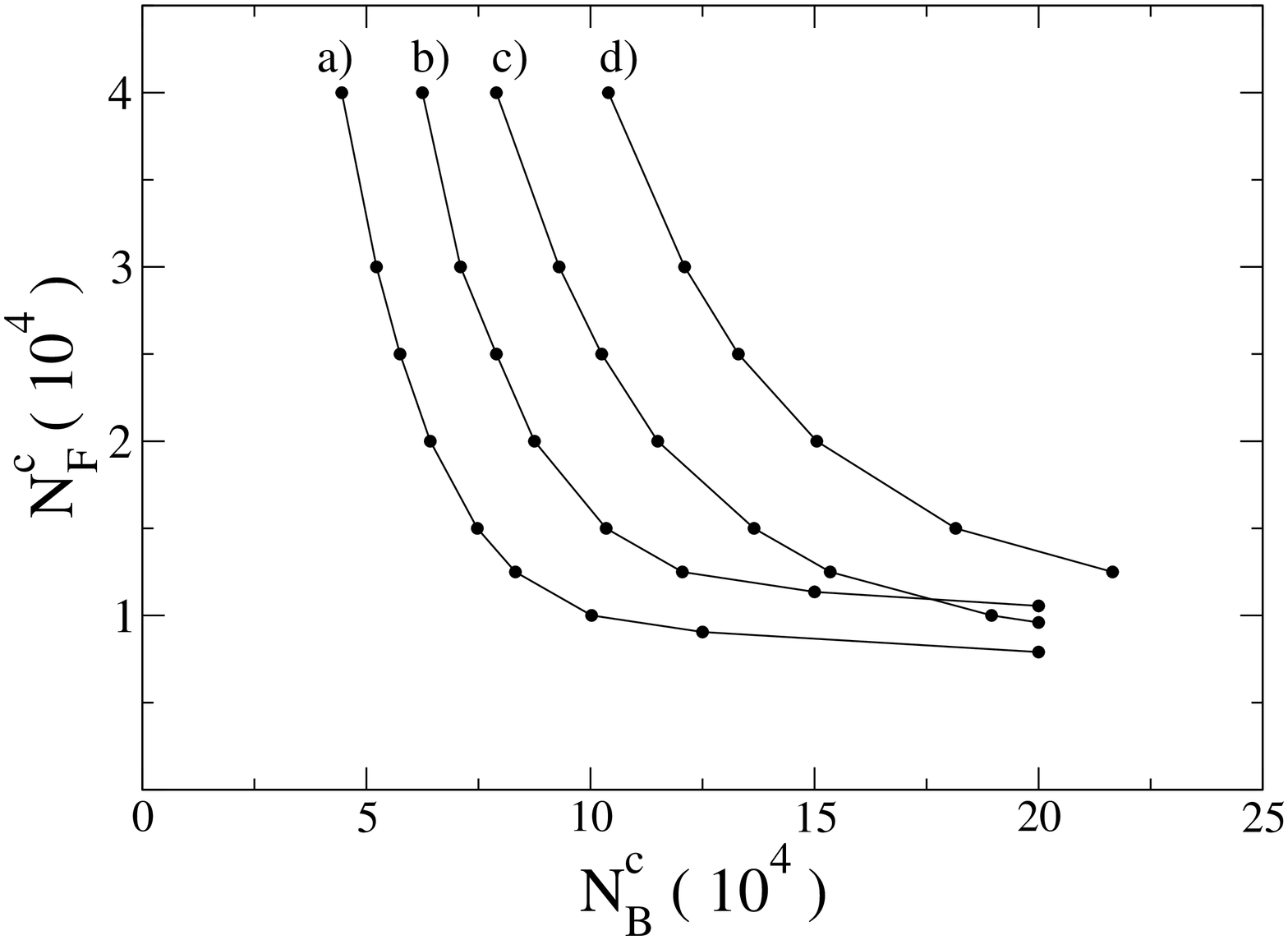}}
\caption[]{
Stability diagram of the $^{40}$K-$^{87}$Rb mixture as a function of
the number of atoms. The dots are the prediction
for the critical number of bosons and fermions.
The lines have been drawn to guide the eye, and represent
the critical instability lines that determine
the boundary between the stable (left) and unstable (right)
regions in four different cases:
a) vortex-free configurations without displacement of the trapping
potentials; b) bosons hosting a vortex line without
displacement of  the trapping
potentials; c) vortex-free configurations with  a $10 \mu m$ 
displacement between the trapping potentials;  d)
bosons hosting a vortex line, plus  a $10 \mu m$
displacement between the
trapping potentials. 
}
\label{fig3}
\end{figure}

We display in Fig. \ref{fig3} the stability diagram for the
$^{40}$K-$^{87}$Rb mixture in the $N_B-N_F$ plane.
The dots are the theoretical prediction for ($N_B^c$,$N_F^c$).
The lines have been drawn to guide the eye and represent
the critical instability lines that determine
the boundary between the stable (left) and unstable (right)
regions in four different cases:
a) vortex-free configurations in concentric trapping potentials
(no displacement of the external fermion and boson potentials); 
b) Bose condensate hosting a vortex line with bosons and fermions confined
by concentric trapping potentials;
c) vortex-free configurations with a $10 \mu m$
displacement between the trapping potentials;  d) Bose condensate 
hosting a vortex line, plus a $10 \mu m$ displacement between the
trapping potentials.

From Fig. \ref{fig3} one can conclude that, for the $^{40}$K-$^{87}$Rb 
mixture, the presence of a vortex in the condensate, or any other mechanism 
that increases the distance between the maxima of the -still- overlapping
densities,
as for example a sag displacement between the two clouds, allows
to have a stable mixture for larger particle numbers.
The reason is that these mechanisms diminish the
enhancement of the density of both species in the overlap volume
caused by their attractive mutual interaction.
A sag displacement (c) leads to
critical numbers higher than the presence of a vortex (b),
and when both are simultaneously present (d), they
yield the larger stability region.

\section{Summary}
We have studied Fermi-Bose mixtures with attractive mutual interaction,
We have
considered two potentially interesting cases. In one of them,
bosons are in a vortex state, and in the other
the minima of the trapping potentials for bosons and fermions are
shifted from each other. 

The position of the maximum of the fermion density is
relevant for the collapse of the trapped boson-fermion mixture. Indeed,
it is around these `fixed'  points that the densities
increase when the atom numbers approach their critical values,
i.e., these are the points where the collapse starts. For this
reason,
simple analytical formulas have been derived to describe the effective
potential felt by a small fermion amount in the presence of a large
boson condensate with positive scattering length. These formulas
have been used to seek the critical points of the fermion density
distribution removing the restriction $N_F << N_B$.
The positions of the critical points only depend on
the value of a dimensionless parameter ($\gamma_i$) we have introduced in
Sec. III.A, whose
definition involves  the values of the boson and fermion scattering
lengths, masses and confining frequencies. 

The validity of these formulas has been assessed comparing the results 
obtained using them with the results obtained from the numerical solution of
the mean-field equations (GP equation for bosons coupled to the TFW
equation for fermions). 
We have numerically shown that the position of the maximum
of the fermionic density is very insensitive to the
dilution value $N_F/N_B$, and
have also checked that the analytical values obtained
for a low number of fermions remain valid up to the critical atomic numbers.

Finally, we have  shown that the critical atomic numbers
that the mixture can sustain before it collapses may be
increased shifting the -still overlapping- confining
potentials, or when the condensate hosts a quantized vortex line.
In the later case, a larger overlap between both species has been
found, which may favor sympathetic cooling.

\section*{Acknowledgements}
This work has been performed under Grants No. BFM2002-01868 from
DGI, Spain, and No. 2001SGR-00064 from Generalitat de Catalunya.
D.M.J. has been also funded by   the
M.E.C.D. (Spain) on sabbatical leave.
 M.G. thanks the `Ram\'on y Cajal'
Programme of the Ministerio de Ciencia y Tecnolog\'{\i}a (Spain)
for financial support.
\appendix
\section*{}

In this Appendix we use a scaling transformation
\cite{pit03,boh79,dal96} to derive
an expression that generalizes the virial theorem and is useful to test
the accuracy of the numerical procedure.
Performing a scaling transformation of the vector position
$\vec{r} \rightarrow \lambda \vec{r}$, it is easy to check that
to keep the normatization of the order parameter $\Psi(\vec{r}\,)$
and of the boson and fermion atomic densities,
they have to transform as:

\begin{eqnarray}
\Psi(\vec{r}\,) \rightarrow
\Psi_{\lambda}(\vec{r}\,)=
\lambda^{3/2} \Psi(\lambda \vec{r}\,)
\nonumber
\\
n(\vec{r}\,) \rightarrow
n_{\lambda}(\vec{r}\,)=
\lambda^3 n(\lambda \vec{r}\,)
\label{a1}
\end{eqnarray}
Splitting the energy of the system into a kinetic
$T$,  a harmonic oscillator $U_H$, and an interaction part $U_g$, the
above transformations induce the following ones:

\begin{eqnarray}
T \rightarrow T_{\lambda} = \frac{\hbar^2}{2 M_B}
\, \int d\vec{r} \,|\nabla \Psi_{\lambda}(\vec{r}\,)|^2
= \lambda^2 T
\nonumber
\\
U_H \rightarrow U_{H_{\lambda}} = \frac{1}{2}\, M\, \omega^2 \,
\int d\vec{r} \, r^2 n_{\lambda} = \frac{1}{\lambda^2} U_H
\nonumber
\\
U_g \rightarrow U_{g_{\lambda}} = \frac{1}{2}\, g \,
\int d\vec{r} \, n^2_{\lambda} = \lambda^3 U_g \;\;\; ,
\label{a2}
\end{eqnarray}
where the last two expressions hold for fermions and bosons as well.
The expression for $U_H$ supposes a spherically symmetric harmonic
trap, the generatization to deformed harmonic traps is obvious.

It is easy to check   that in the case of fermions, the kinetic energy
in the TFW approximation also scales as
$T \rightarrow T_{\lambda} = \lambda^2 T$, and
that in the case of boson-fermion mixtures, the  interaction term
$U_{g_{BF}}$ scales as
 $U_{g_{BF}} \rightarrow U_{g_{{BF}_{\lambda}}} = \lambda^3 U_{g_{BF}}$.
Hence, the total energy of the mixture scales as

\begin{equation}
E \rightarrow E_{\lambda}= \lambda^2 (T_B + T_F)
 +\frac{1}{\lambda^2} (U_{H_B} + U_{H_F} ) +  \lambda^3 ( U_{g_{BB}} +
U_{g_{FF}} + U_{g_{BF}}) \;\;\; .
\label{a3}
\end{equation}

If the scaling is carried out from the equilibrium configuration
corresponding to the value $\lambda=1$, one has

\begin{equation}
\left.\frac{d E_{\lambda}}{d \lambda}\right|_{\lambda=1} =
2 (T_B + T_F) - 2 (U_{H_B} + U_{H_F} ) + 3  ( U_{g_{BB}} + U_{g_{FF}} +
U_{g_{BF}}) = 0  \;\;\; .
\label{a4}
\end{equation}
One sees that at equilibrium, for non-interacting fermions and bosons
one recovers the virial theorem, namely $ T_B = U_{H_B}$ and $T_F = U_{H_F}$.

When a vortex is present, its superfluid kinetic energy
\begin{equation}
T_V  = \frac{\hbar^2}{2 M_B}  \,
\int d\vec{r} \; \frac{n_B(\vec{r}\,)}{r_{\perp}^2}
\label{a5}
\end{equation}
scales as
\begin{equation}
T_V \rightarrow T_{V_{\lambda}} = \frac{\hbar^2}{2 M_B}
\, \int d\vec{r} \, \frac{n_{\lambda}(\vec{r}\,)}{r_{\perp}^2}
 = \lambda^2 T_V
\label{a6}
\end{equation}
This is quite a natural result in view of the first Eq. (\ref{a2})
and, as a consequence, $T_V$ may be incorporated in the definition of
$T_B$ which then represents the total kinetic energy of the
bosonic component of the mixture.

If a gravitational sag is considered that displaces
the atomic clouds in the $z$ direction, its effect can be
taken into account changing the confining potential in that
direction into $V(z)= \frac{1}{2} M\, \omega^2 (z -z_0)^2$.
Apart from a trivial constant term proportional to the number
of fermions and bosons in the trap that is invariant under the
scaling transformation, there appear new contributions to the
total energy of the kind

\begin{equation}
U_s  = - M z_0 \,\omega^2  \,
\int d\vec{r} \, z \, n(\vec{r}\,)
\label{a7}
\end{equation}
which scales as
$U_s \rightarrow U_{s_{\lambda}} = \frac{1}{\lambda} U_s$.
Eq. (\ref{a4}) then becomes

\begin{equation}
\left.\frac{d E_{\lambda}}{d \lambda}\right|_{\lambda=1} =
2 (T_B + T_F) - 2 (U_{H_B} + U_{H_F} ) -(U_{s_B} + U_{s_F})
+ 3 ( U_{g_{BB}} + U_{g_{FF}} + U_{g_{BF}}) = 0
\;\;\; .
\label{a8}
\end{equation}
The above expression is, in a way, a generalization of the virial
theorem. We have used it to routinely check the  accuracy
in the numerical solution of the GP and
TFW coupled equations. Values $\sim O(10)$ are
found for that expression when $U_H$ is
$\sim O(10^6)$. This makes us confident on the numerical method
we have used to find the mean-field structure of the atomic mixture.


%

\begin{thebibliography}{99}

\bibitem{tr01} A. G. Truscott, K. E. Strecker, W. I.  McAlexander,
G. B. Partridge, and R. G. Hulet, Science {\bf 291}, 2570 (2001).

\bibitem{sc01} F. Schreck, L. Khaykovich, K. L. Corwin, G. Ferrari,
T.  Bourdel, J. Cubizolles, and C. Salomon, Phys. Rev. Lett.
{\bf 87}, 080403  (2001).

\bibitem{ro02} R. Roth, Phys. Rev. A {\bf 66}, 013614 (2002).

\bibitem{ak02} Z. Akdeniz, P. Vignolo, A. Minguzzi, and M. P. Tossi,
J. Phys. B {\bf 35}, L105 (2002).

\bibitem{fe02}
G. Ferrari, M. Inguscio, W. Jastrzebski, G. Modugno, and G. Roati,
Phys. Rev. Lett. {\bf 89}, 053202 (2002).

\bibitem{roa02}
G. Roati, F. Riboli, G. Modugno, and M. Inguscio,
Phys. Rev. Lett. {\bf 89}, 150403 (2002).

\bibitem{mo03}
M. Modugno, F. Ferlaino, F. Riboli, G. Roati, G. Modugno, and M.
Inguscio, 
Phys. Rev. A {\bf 68}, 043626 (2003).

\bibitem{mo98}
 K. Molmer, Phys. Rev. Lett. {\bf 80}, 1804 (1998).

\bibitem{vi00}
 L. Vichi, M. Amoruso, A. Minguzzi, S. Stringari, and M. Tosi
Eur. Phys. J. D {\bf 11}, 335 (2000).

\bibitem{don91}
R. J. Donnelly, {\it Quantized Vortices in Helium II}
(Cambridge University Press, Cambridge, 1991).

\bibitem{mat99}
M. R. Matthews, B. P. Anderson, P. C. Haljan, D. S. Hall, C.E.
Wieman, and E. A. Cornell, Phys. Rev. Lett. {\bf 83}, 2498 (1999).

\bibitem{fet01} A. L. Fetter and A. A. Svidzinsky,
J. Phys.: Condens. Matter {\bf 13}, R135 (2001).

\bibitem{pet02}
C. J. Pethick and H. Smith, {\it Bose-Einstein  Condensation in Dilute
Gases} (Cambridge University Press, Cambridge, 2002).

\bibitem{pit03}
L. Pitaevskii and S. Stringari, {\it Bose-Einstein Condensation}
(Clarendon Press, Oxford, 2003).

\bibitem{may01}
R. Mayol, M. Pi, M. Barranco, and F. Dalfovo, Phys. Rev. Lett.
{\bf 87}, 145301 (2001).

\bibitem{ca03}
P. Capuzzi, A. Minguzzi, and M.P. Tosi,
Phys. Rev. A {\bf 68}, 033605 (2003).

\bibitem{pi88}
M. Pi, X. Vi\~nas, F. Garcias, and M. Barranco,
Phys. Lett. B {\bf 215}, 5 (1988).

\bibitem{bar03}
M. Barranco, M. Guilleumas, E.S. Hern\'andez, R. Mayol, M. Pi, and L.
Szybisz, Phys. Rev. B {\bf 68}, 024515 (2003).

\bibitem{ca01}
P. Capuzzi and E.S. Hern\'andez, Phys. Rev. A {\bf 64}, 043607 (2001).

\bibitem{boh79}
O. Bohigas, A.M. Lane, and J. Martorell, Phys. Rep. {\bf 51}, 267 (1979).

\bibitem{dal96}
F. Dalfovo  and S. Stringari, Phys. Rev. A {\bf 53}, 2477 (1996).


\end{thebibliography}
\end{document}